\documentclass[pdflatex,sn-mathphys-num]{sn-jnl}
\usepackage{graphicx}%
\usepackage{multirow}%
\usepackage{amsmath,amssymb,amsfonts}%
\usepackage{amsthm}%
\usepackage{mathrsfs}%
\usepackage[title]{appendix}%
\usepackage{textcomp}%
\usepackage{manyfoot}%
\usepackage{booktabs}%
\usepackage{algorithm}%
\usepackage{algorithmicx}%
\usepackage{algpseudocode}%
\usepackage{listings}%

\usepackage{mathtools}
\usepackage{braket}
\usepackage{slashed} 
\usepackage{physics}
\usepackage{bbding}
\usepackage{bbold}

\usepackage{hyperref} 

\renewcommand{\footnoterule}{%
  \kern -3pt                         % Remonte de 3pt pour compenser l'épaisseur
  \hrule width 0.5\textwidth          % Trace une ligne de 50% de la largeur du texte
  \kern 2.6pt                        % Redescend pour laisser un espace avant la note
}

\usepackage{color}

\newcommand{\cba}{c_{\beta-\alpha}}
\newcommand{\sba}{s_{\beta-\alpha}}

\begin{document}

\title[Article Title]{Possibility of Probing an Extra Higgs Boson at the Compact Linear Collider}


\author{\fnm{Mohamed} \sur{Krab}\footnote{E-mail: mkrab@hep1.phys.ntu.edu.tw}} 
%\email{mkrab@hep1.phys.ntu.edu.tw}

\affil{\orgdiv{Department of Physics}, \orgname{National Taiwan University}, \orgaddress{\street{No. 1, Sec. 4, Roosevelt Rd.}, \city{Taipei}, \postcode{10617}, \country{Taiwan}}}


\abstract{We study the sensitivity of the Compact Linear Collider (CLIC) to an additional neutral Higgs boson $H$ through the vector boson fusion process $e^+e^- \to H\nu\bar \nu$, followed by the decay $H \to W^+W^-$, with both $W$ bosons decaying leptonically, resulting in a dilepton plus missing transverse energy final state.
Within the framework of the Two Higgs Doublet Model, where both production and decay are governed by the Higgs mixing angle $\cos(\beta-\alpha)$, we perform a detector-level analysis and show that a high-energy CLIC can probe $H$ via this channel, allowing a direct measurement of the $HWW$ coupling.}

%%================================%%
%% Sample for structured abstract %%
%%================================%%

\keywords{Beyond Standard Model, Higgs Physics, Compact Linear Collider}



\maketitle
\newpage

\section{Introduction}
The discovery of a 125 GeV neutral Higgs boson provides strong evidence for the Higgs mechanism of Electroweak Symmetry Breaking (EWSB) within the Standard Model (SM), reinforcing its validity at currently accessible collider energies. Yet, the SM remains incomplete, as it cannot explain several observed phenomena, including the Baryon Asymmetry of the Universe (BAU). Electroweak Baryogenesis (EWBG), a mechanism proposed to explain the BAU, requires a strong first-order electroweak phase transition and additional sources of CP violation beyond the SM. Extended Higgs sectors, such as the Two Higgs Doublet Model (2HDM), provide a natural framework to satisfy these requirements, potentially offering a viable explanation for the observed BAU. 
The discovery of additional Higgs bosons would unambiguously signal an extended scalar sector.

The 2HDM is among the simplest and well-motivated extensions of the SM Higgs sector, featuring an additional scalar doublet. After EWSB, the scalar spectrum of the CP-conserving 2HDM consists of two CP-even neutral scalars, $h$ and $H$, one CP-odd neutral scalar (the pseudoscalar $A$), and a pair of charged Higgs bosons, $H^\pm$.
%Here, we adopt the standard mass hierarchy in which the observed Higgs boson with a mass of 125 GeV is identified with the lighter CP-even state $h$.
In its most general realization, the 2HDM allows both Higgs doublets to couple to all fermions, which generically leads to tree-level flavor-changing neutral currents (FCNCs). 
This realization is often referred to as type-III 2HDM or General 2HDM (G2HDM). Such a model allows for a second set of Yukawa couplings that could carry CP-violating phases and hence drive EWBG (see, e.g., Refs.~\cite{Fuyuto:2017ewj,Modak:2018csw,Aiko:2025tbk}).  
The G2HDM allows flavor-changing neutral Higgs (FCNH) decays, such as $t \to ch$~\cite{Hou:1991un,Chen:2013qta}. 
Recent searches for $t \to ch$ by ATLAS~\cite{ATLAS:2024mih} and CMS~\cite{CMS:2024ubt} have set stringent limits on the branching ratio 
$\mathcal{B}(t \to ch) < 3.3\times 10^{-4}$~\cite{ATLAS:2024mih} and $\mathcal{B}(t \to ch) < 3.7\times 10^{-4}$~\cite{CMS:2024ubt}, respectively. (Interpretation of these limits in terms of G2HDM can be found in Ref.~\cite{Krab:2025zuy}.)
Given the large expected yield of top quarks and relatively low background levels, Compact Linear Collider (CLIC)~\cite{Linssen:2012hp,CLIC:2016zwp,Aicheler:2018arh} provides competitive searches
for $t \to ch$ decay, where an expected upper limit of $\mathcal{B}(t \to ch) \times \mathcal{B}(h \to b\bar b) < 8.8\times 10^{-5}$~\cite{CLICdp:2018esa} can be achieved with an integrated luminosity of $1.0~\rm{ab}^{-1}$ and center-of-mass energy $\sqrt{s} = 380$~GeV.
However, FCNCs can be avoided by imposing a discrete $Z_2$ symmetry, yielding the so-called 2HDM with natural flavor conservation~\cite{Branco:2011iw}.

Among proposed future colliders, CLIC is a multi-TeV $e^+e^-$ machine that combines the high-energy reach with the clean experimental environment characteristic of lepton colliders. CLIC is planned to operate in three stages with center-of-mass energies $\sqrt{s} = 380$~GeV, 1.5~TeV, and 3~TeV, providing an exceptional setting for precision studies of the Higgs sector.
In addition, the 1.5 and 3~TeV high-energy stages enable direct probes of heavy Higgs bosons, which are complementary to precision measurements of the properties of the 125 GeV Higgs boson and may offer a promising avenue for exploring an extended Higgs sector.

In this work, we investigate the prospects for probing an additional Higgs boson at CLIC.  
We focus on the production process $e^+ e^- \to H\nu \bar\nu$, followed by the decay $H \to WW$. Assuming leptonic decays of the $W$ bosons,\footnote{Final states with hadronic $W$ decays are discussed in Ref.~\cite{Hou:2025fiy}.} the signal is characterized by a pair of opposite-sign leptons accompanied by missing transverse energy. We analyze the relevant SM backgrounds and assess the discovery potential at $\sqrt{s} = 1.5$ and 3~TeV, demonstrating that CLIC can achieve sensitivity to the Heavy Higgs boson $H$ via $e^+e^- \to H\nu\bar\nu \to W^+W^-\nu\nu$.

\section{The model}
In the Higgs basis, one can write the most general CP-conserving scalar potential of the G2HDM as~\cite{Davidson:2005cw,Hou:2017hiw}
\begin{align}
V(\Phi,\Phi') =& ~\mu_{11}^2|\Phi|^2 + \mu_{22}^2|\Phi'|^2
    - (\mu_{12}^2\Phi^\dagger\Phi' + \rm{H.c.}) \nonumber\\
 & + \frac{\eta_1}{2}|\Phi|^4 + \frac{\eta_2}{2}|\Phi'|^4
   + \eta_3|\Phi|^2|\Phi'|^2  + \eta_4 |\Phi^\dagger\Phi'|^2 \nonumber \\
 & + \left[\frac{\eta_5}{2}(\Phi^\dagger\Phi')^2
   + \left(\eta_6 |\Phi|^2 + \eta_7|\Phi'|^2\right) \Phi^\dagger\Phi' + \rm{H.c.}\right],
\label{pot}
\end{align}
where $\eta_{1\text{--}7}$ are the Higgs quartic couplings and are real, 
$\Phi$ generates the vacuum expectation value $\mathit{v}$ for EWSB, with minimization condition $\mu^2_{11} = -\frac{1}{2} \eta_1 v^2$, while $\left\langle \Phi'\right\rangle = 0$ hence $\mu_{22}^2 > 0$. The second condition $\mu^2_{12} = \frac{1}{2}\eta_6 v^2$ reduces the parameter count by one.
The mixing angle $\beta-\alpha$ diagonalizes the CP-even Higgs squared-mass matrix~\cite{Haber:2015pua}. The squared masses of $h$ and $H$ are given by
\begin{align}
  m_{h,H}^2 = \frac{1}{2}\left[m_A^2 + (\eta_1 + \eta_5) v^2 \mp \sqrt{\left(m_A^2+ (\eta_5 - \eta_1) v^2\right)^2
   + 4 \eta_6^2 v^4}\right].
\end{align}

The Yukawa couplings are given by \cite{Davidson:2005cw}
\begin{align}
\mathcal{L}_Y = %\supset
& - \frac{1}{\sqrt{2}} \sum_{f = u, d, \ell}
\bar f_{i} \left[\big(\lambda^f_{ij} \sba + \rho^f_{ij} \cba \big)h 
 + \big(\lambda^f_{ij} \cba - \rho^f_{ij} \sba \big) H \right. \nonumber \\
& \left. \qquad\qquad\qquad\quad -i\,{\rm sgn}(Q_f) \rho^f_{ij} A \right]  P_R f_{j}
- \bar{u}_i\big[(V\rho^d)_{ij} P_R - (\rho^{u\dagger}V)_{ij} P_L \big]d_j H^+ \nonumber \\
&  \qquad\qquad\qquad\quad - \bar{\nu}_i\rho^\ell_{ij} P_R \ell_j H^+ +{\rm H.c.},
\label{LYukawa}
\end{align}
where $i,j = 1\text{--}3$ are the generation indices, $\cba \equiv \cos(\beta-\alpha)$ is the $h\text{--}H$ mixing angle, $\sba \equiv \sin(\beta-\alpha)$, $P_{L,R} = (1\mp\gamma_5)/2$,
and $V$ denotes the CKM matrix. 
The matrices $\lambda^f_{ij}$ are real, diagonal, and given by $\lambda^f_{ij} \equiv \delta_{ij}\sqrt{2}m_i^f/v$, while the matrices $\rho^f_{ij}$ are, in general,
complex and non-diagonal. 
In what follows, we drop the superscript $f$ and assume $\rho_{ij}$ to be real.

%For bibliography use \cite{RefJ}, \cite{RefB}
\section{Benchmark Scenario}
\begin{figure}[t]
	\centering
	\includegraphics[scale=0.4]{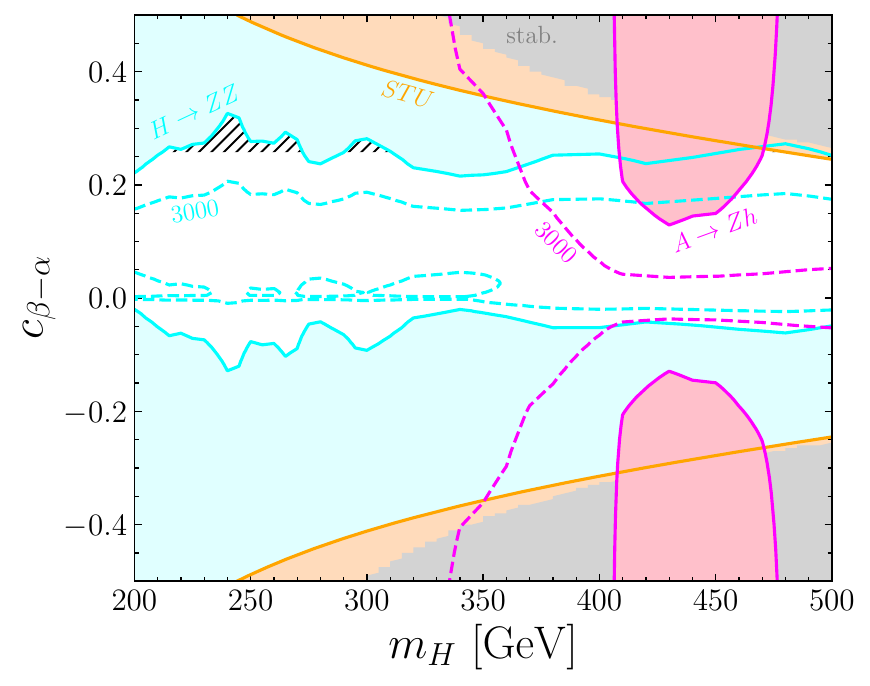}	
	\includegraphics[scale=0.4]{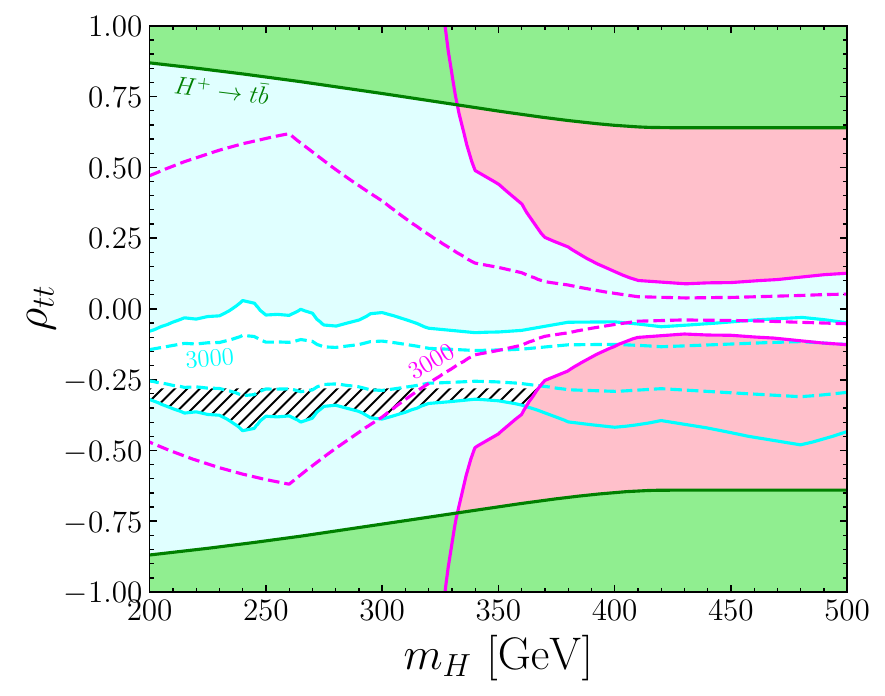}									
	\caption{Exclusion contours in the $m_H$-$\cba$ plane for $\rho_{tt} = -0.1$ (left) and in the $m_H$-$\rho_{tt}$ plane for $\cba = 0.2$ (right). Shown are constraints from vacuum stability (light gray), oblique parameters $S$, $T$, and $U$ (orange), SM Higgs signal strengths (hatched), as well as limits from $H \to ZZ$ (cyan), $A \to Zh$ (magenta), and $H^+ \to t\bar b$ (green) searches. The expected HL-LHC search sensitivities for $H \to ZZ$ (dashed cyan) and $A \to Zh$ (dashed magenta) are also shown.}	
	\label{fig:HB}
\end{figure}

We consider G2HDM with conservative extra top Yukawa coupling ($|\rho_{tt}| = 0.1 $), though still robust in driving EWBG~\cite{Fuyuto:2017ewj}.
We assume $m_A = m_{H^+} = 500$~GeV and scan $\cba \in [-0.5, 0.5]$, $\rho_{tt} \in [-1,1]$, $m_H \in [200, 500]~$GeV. The lighter CP-even scalar $h$ is identified with the observed SM-like Higgs boson, with its mass set to $m_h = 125$~GeV.
The remaining parameters are chosen to be $\mu^2_{22}=m^2_H / 2$, $\eta_2 \simeq 2.52$ and $\eta_7 \simeq 0.17$ to satisfy the theoretical constraints.
In Fig.~\ref{fig:HB} (left panel), we show the exclusion contours in the $m_H$-$\cba$ plane, assuming $\rho_{tt} = -0.1$, where the most sensitive limits come from searches for heavy resonances in the $H \to ZZ$~\cite{ATLAS:2020tlo} (cyan) and $A \to Zh$~\cite{ATLAS:2022enb} (magenta) channels. These exclusion limits are obtained with \textsc{HiggsBounds}\footnote{The ATLAS $A \to Zh$ search limit~\cite{ATLAS:2022enb} is applied by hand. CMS limit~\cite{CMS:2024vxt} is weaker and thus not shown.} module of \textsc{HiggsTools}~\cite{Bahl:2022igd}.
Hatched region is excluded by SM Higgs signal strength measurements, as obtained with \textsc{HiggsSignals} module of \textsc{HiggsTools}.
The orange region is excluded by constraints from oblique parameters $S$, $T$ and $U$, while the gray region is excluded by theoretical requirements of vacuum stability.  
Assuming $\cba = 0.2$, we plot the exclusion contours in the $m_H$-$\rho_{tt}$ plane in the right panel of Fig.~\ref {fig:HB}. 
As in the left panel, for small values of $\rho_{tt}$ the most stringent constraints arise from $H \to ZZ$~\cite{ATLAS:2020tlo} (cyan) and $A \to Zh$~\cite{ATLAS:2022enb} (magenta).
At larger $\rho_{tt}$ coupling, additional sensitivity is provided by the channels $H^+ \to t\bar b$~\cite{ATLAS:2021upq} (green) and $A \to t\bar t$~\cite{CMS:2019pzc}, once the $t\bar t$ decay mode becomes kinematically accessible. 
The constraints from $H^+ \to W^+h$~\cite{ATLAS:2024rcu} searches are very weak.
The expected sensitivity of the high-luminosity LHC (HL-LHC), obtained using a $\sqrt{\mathcal{L}}$ extrapolation, is also shown in Fig.~\ref{fig:HB}.

\section{Collider Study}
\begin{figure}[t]
	\centering
	\includegraphics[scale=0.42]{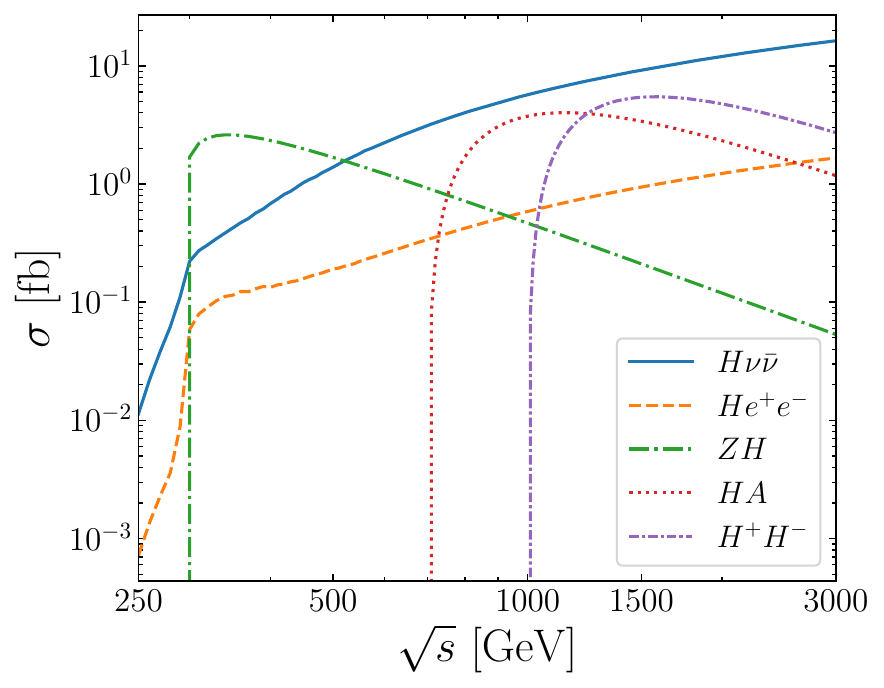}									
	\caption{Leading order cross sections vs. $\sqrt{s}$ for $m_{H} = 200$~GeV, $m_A = m_{H^+} = 500$~GeV, and $\cba = 0.2$. The cross sections are obtained with \textsc{MadGraph5\_aMC@NLO}~\cite{Alwall:2014hca}.}	
	\label{fig:CSs}
\end{figure}

We investigate the discovery potential of the additional neutral Higgs boson $H$ at CLIC, assuming $\sqrt{s} = 1.5$~TeV and an integrated luminosity of 2.5~ab$^{-1}$~\cite{CLICdp:2018esa}. The signal process is $e^+e^- \to H \nu \bar\nu \to W^+ W^- \nu \bar\nu$, where both $W$ bosons decay leptonically, leading to the final state $\ell^+\ell^-\nu\bar\nu \nu\bar\nu$ (hereafter denoted as $\ell\ell + E^{\rm miss}_T$).
We choose $m_H = 200~(250)$~GeV, $m_A = m_{H^+} = 500$~GeV, $\cba = 0.2$, and $\rho_{tt} = -0.1$ as a representative benchmark point, denoted BP1 (BP2), for which the predominant decay channel of $H$ is $H \to W^+ W^-$, followed by $H \to ZZ$.
%\footnote{We do not include the $H \to ZZ$ channel, though it would lead to the same final state via $Z \to \ell^+ \ell^-$ and $Z \to \nu\bar\nu$.}
In our study, only $WW$-fusion is considered for the $H$ production. Production through $ZZ$-fusion, i.e. $e^+ e^- \to H e^+e^-$, as well as $e^+e^- \to ZH$ and $e^+ e^- \to HA$ processes exhibit relatively small production cross sections at TeV energies (see Fig.~\ref{fig:CSs}).
The main SM backgrounds come from $\ell^+\ell^-$ (referred to as $Z/\gamma^*$), $WW$, $ZZ$, $t\bar t$, and $h\nu \bar\nu$ ($h$ refers to the SM-like Higgs boson). 
Other backgrounds, such as $Zh$ and $he^+e^-$ are minor.

Signal and background events are generated at leading order using  \textsc{MadGraph5\_aMC@NLO}~\cite{Alwall:2014hca}, then passed to \textsc{Pythia-8.2}~\cite{Sjostrand:2014zea} for parton showering and hadronization, and to \textsc{Delphes-3.5.0} \cite{deFavereau:2013fsa}, for fast detector simulation, with the default \texttt{CLICdet\_Stage2} card~\cite{CLICdp:2018vnx}, where the jet clustering is performed using the Valencia algorithm~\cite{Boronat:2014hva,Boronat:2016tgd} via \textsc{FastJet}~\cite{Cacciari:2011ma}. 
Throughout this analysis, the incoming electron and positron beams are assumed to be unpolarized.
Including beam polarisation, however, would enhance the $e^+e^- \to H\nu\bar\nu$ production rate~\cite{Abramowicz:2016zbo}.

\begin{table*}[b]
	\centering
	{\small \begin{tabular}{l c c c c c c c} %| c| c 
			\hline\hline
			$\sqrt{s} = 1.5$~TeV & BP1 & BP2 & $t\bar t$ & $WW$ & $ZZ$ & $Z/\gamma^*$ & $h\nu \bar\nu$ \\\hline
            Exactly two OS leptons & 0.189 & 0.160 & 3.176 & 25.289 & 3.200 & 11013.7 & 1.089 \\		
            $120 < m^{\ell\ell}_T < 260$ GeV  & 0.124 & 0.104 & 0.469 & 1.362 & 0.669 & 0.304 & 0.618 \\	
            $E^{\rm{miss}}_T > 50$ GeV        & 0.104 & 0.074 & 0.373 & 0.295 & 0.055 & 0.120 & 0.569 \\			
            Jet veto                          & 0.104 & 0.074 & $<0.01$ & 0.284 & 0.023 & 0.120 & 0.555 \\ 
            $p^{\ell\ell}_T > 30$ GeV         & 0.104 & 0.074 & $<0.01$ & 0.263 & 0.023 & 0.101 & 0.553 \\ \hline\hline							
	\end{tabular}}
	\caption{Signal and background cross sections (in fb) at each event selection cut.} 
	\label{table:xsASC} %ttbar after jet veto 0.0009
\end{table*} 

\begin{figure*}[t]
	\centering
	\includegraphics[scale=0.4]{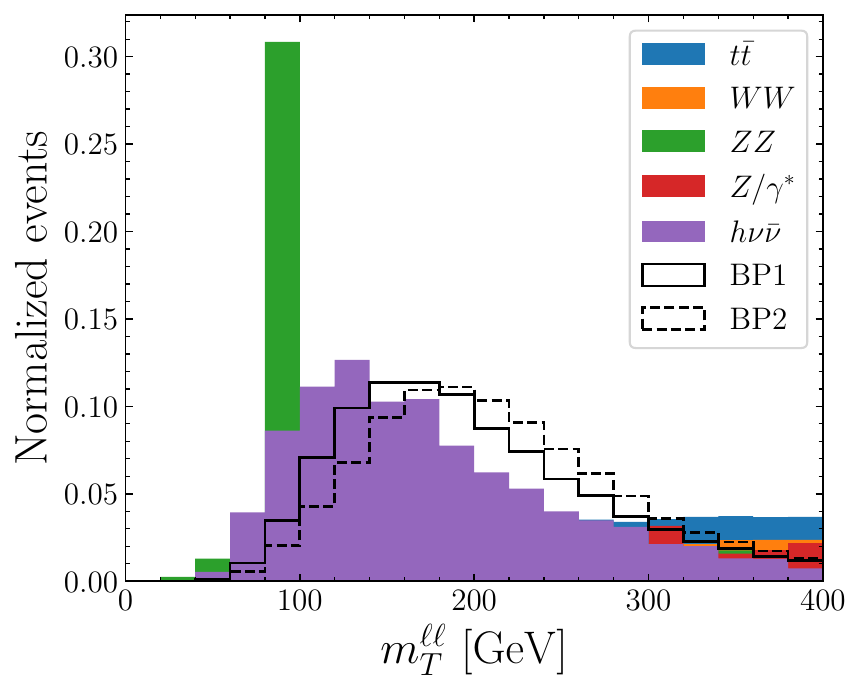}
	\includegraphics[scale=0.4]{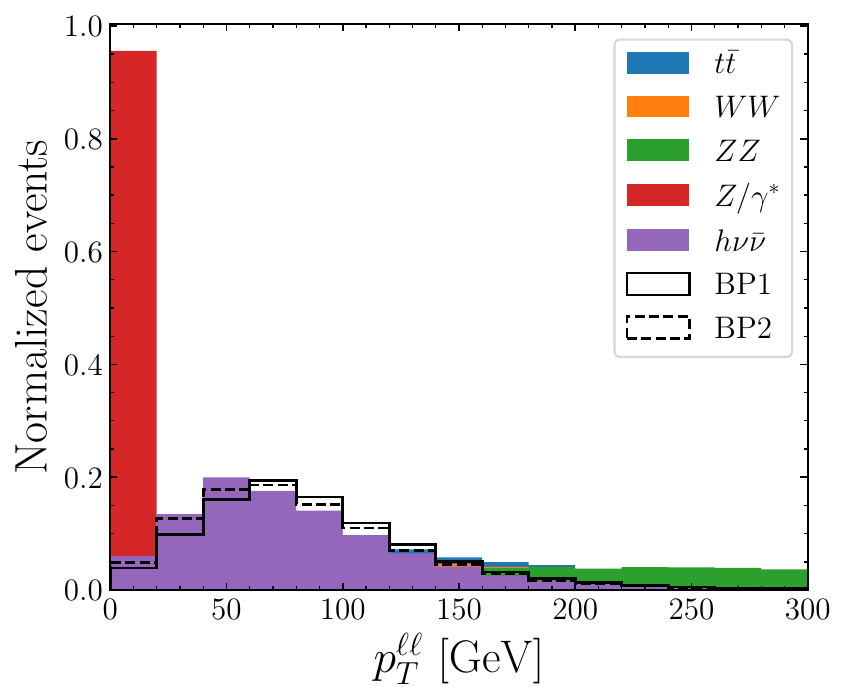}			
	\caption{Normalized distributions for signal and background processes at 1.5~TeV CLIC.}	
	\label{fig:distr}
\end{figure*}

Events are required to contain exactly two opposite-sign (OS) leptons with transverse momentum $p_T > 20$ and $|\eta|<2.5$. After this preselection, the event sample is dominated by the $Z/\gamma^*$ background. To reduce this contribution, a selection cut on the dilepton transverse mass, $m^{\ell\ell}_T$, is imposed.
This requirement is also effective in reducing backgrounds from $t\bar t$, $WW$, and $ZZ$ production.
The missing transverse energy, $E^{\rm miss}_T$, is subsequently used to further suppress the $ZZ$ background.
After applying the $m^{\ell\ell}_T$ and $E^{\rm miss}_T$ selection cuts, the remaining background is dominated by $h\nu\bar\nu$, $t\bar t$, and $WW$ processes. To suppress the $t\bar t$ contribution, events containing at least one jet candidate with $p_T > 25$~GeV are vetoed (referred to as ``jet veto'').
At this stage, the signal-to-background ratio is already significantly improved.
However, to further reduce the $Z/\gamma^*$ background, the transverse momentum of the dilepton system is required to satisfy $p^{\ell\ell}_T > 30$~GeV.
All selection criteria are optimized to maximize the statistical significance of the signal.
The cross sections of the simulated signal and background processes after each selection cut are summarized in Table~\ref{table:xsASC}.
The $m^{\ell\ell}_T$ and $p^{\ell\ell}_T$ distributions are shown in Fig.~\ref{fig:distr}.

\begin{figure*}[b]
	\centering
	\includegraphics[scale=0.4]{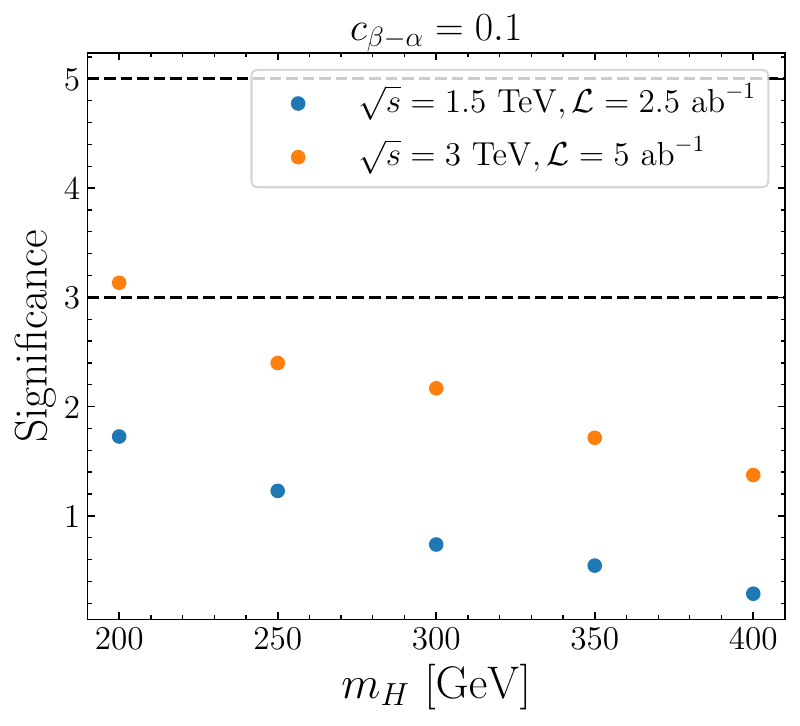}	
	\includegraphics[scale=0.4]{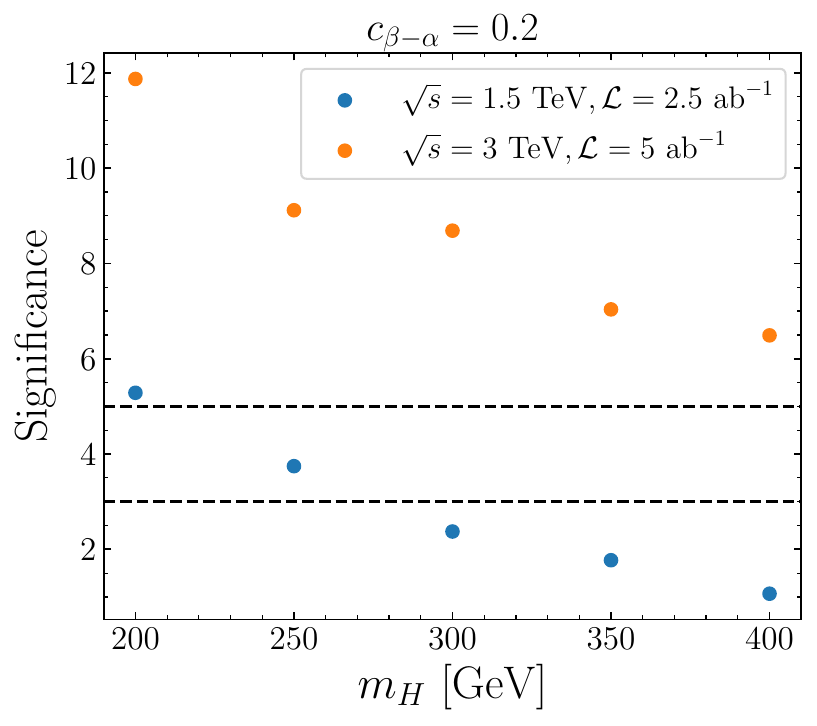}									
	\caption{Significance as a function of $m_H$ for 1.5 and 3~TeV CLIC.}	
	\label{fig:significance}
\end{figure*}

We now estimate the signal significance using the cross sections summarized in Table~\ref{table:xsASC}. The significance is computed using~\cite{Cowan:2010js}
\begin{equation}
    \mathcal{Z} = \sqrt{2\left[\left(N_S+N_B\right)\ln\left(1+\frac{N_S}{N_B}\right)-N_S\right]},
\end{equation}
where $N_S$ and $N_B$ denote the number of signal and background events, respectively. 
Assuming an integrated luminosity of 2.5~ab$^{-1}$, BP1 (BP2) yields a significance of approximately $5.3\sigma$ (3.7$\sigma$).
Applying the same selection criteria as in Table~\ref{table:xsASC}, with exception that the dilepton transverse mass is restricted to $140 < m^{\ell\ell}_T < 340$~GeV, we obtain $\mathcal{Z} \simeq 2.4\sigma$ for $m_H = 300$~GeV. 
Significance as a function of $m_H$ for $\sqrt{s} = 1.5$ and 3~TeV is presented in Fig.~\ref{fig:significance}.

\section{Conclusion}
We considered a benchmark scenario with $\cba \neq 0$ and small extra top Yukawa coupling $\rho_{tt}$, while setting the FCNH coupling $\rho_{tc}$ to zero. Allowing a nonzero $\rho_{tc}$ would induce $H \to t\bar c$ and soften $H \to WW$, although the latter remains predominant.
For $\rho_{tc} = 0.1$, the significance for BP1 is reduced to approximately $4.6\sigma$ at $\sqrt{s} = 1.5$~TeV and $\mathcal{L} = 2.5~\rm{ab}^{-1}$.
The additional states $A$ and $H^+$ can be searched for through $e^+e^- \to HA \to H ZH, HZh$ and $e^+e^- \to H^+H^- \to W^+H W^-H, W^+hW^-h$, respectively, but this would require further investigation.

In summary, we have investigated the prospects for probing an extended Higgs sector through the production and decay of a heavy neutral Higgs boson in a clean lepton-collider environment. We have shown that high-energy CLIC can probe an additional Higgs boson through the process $e^+e^- \to H \nu \bar\nu \to W^+ W^- \nu \bar\nu$. This channel can also provide a direct probe of the $HWW$ coupling.

\section*{Acknowledgments}
I thank George Wei-Shu Hou for collaboration.
This work was supported by the NSTC of Taiwan under grant No.~114-2639-M-002-006-ASP. 

%\section*{Data Availability Statement}
%Data sets generated during the current study are available from the corresponding author on reasonable request.

\bibliography{proceedings-LCWS}

\end{document}